# Electron Delocalization Determined Anomalous Stability in Small Water Rings

Bo Wang[1], Minsi Xin[1], Xing Dai[1], Ruixia Song[1], Yan Meng[1], Zhigang Wang[1,*] and Ruiqin Zhang[2,3,*]

[1]Institute of Atomic and Molecular Physics, Jilin University, Changchun 130012, China

[2]Department of Physics and Materials Science and Centre for Functional Photonics (CFP), City University of Hong Kong, Hong Kong SAR, China

[3]Beijing Computational Science Research Center, Beijing 100084, China.

*To whom correspondence should be addressed. E-mail: wangzg@jlu.edu.cn, Tel.: +86-431-85168817; aprqz@cityu.edu.hk, Tel.: +852-34427849.

**Abstract**: Water clusters are known to form through hydrogen bonding. However, this study shows that the formation of very small water clusters significantly deviates from this mechanism and instead involves both hydrogen bonding and electron delocalization. Our density functional theory calculations show that small water rings $(H_2O)_n$ of n=3 or 4 show strong electron delocalization originating from both the hydrogen and oxygen atoms and extending to the ring center. This is very different from larger rings. Further energy decomposition of rings with n=3-6 demonstrates an upward trend in the polarization component but an decrease in the electrostatic and exchange repulsion components, presenting a minimum and accounting for 33% of interaction energy at n=3. This significantly promotes stability of the small water rings. Our findings provide a comprehensive analysis and improve our understanding of the stability characteristics of water clusters.

**Introduction**

Water clusters are the elementary structures and functional units of water. For a long time, there has been a basic consensus that the origin of the formation of water clusters is hydrogen bonding, the simplest form of electrostatic attraction[1-6]. Studies confirm that stable water cluster structures are formed through complex hydrogen bond connections[7,8]. Based on this understanding, numerous complicated structures including water rings, and even the functional characteristics of water clusters, have been elucidated. Among water rings, the isolate single water ring (ISWR) is a typical closed system with n hydrogen bonds formed by n water molecules[1, 9]. ISWR structures for n=3-5[10,11] have been found to exist in experiments using far-infrared vibration-rotation tunneling (FIR-VRT) spectroscopy[10] and infrared (IR) laser spectroscopic technologies[12], while for n = 6, the ISWR structure reaches the maximum value of the water ring and tends to change into the cage type with varied geometric structures[13,14]. Quantum-mechanical level theoretical calculations have been used to study the conformations[8,15,16], energy[17,18], IR or Raman spectroscopy[2], intramolecular vibrational redistribution and vibrational energy transfer[19], confined water structure and proton transport process[16,20], and so on of water clusters. In particular, after differences emerged between the results of adiabatic potential energy surface and experimental studies[13], there has been a closer collaboration between theoretical and experimental researchers.

So far, the belief is that for clusters with n<6, the cyclic structure has the lowest energy[1,9,11,15,21]. However, this may not represent a complete understanding because for cyclic structures, although a closed structure is formed, the curvature of the circle for n=3 is larger than that for n=6 for the cases with n=3-6, leading to an over-bent hydrogen bond angle[9,21]. Is electrostatic attraction alone enough to stabilize such a cyclic structure? To answer this question, we have carried out a detailed analysis of the electronic structure properties of cyclic structures of n=3-6, with a focus on the mechanism of the anomalous stability of water rings at n=3 and 4. Our study was conducted using density functional theory (DFT) methods and based on analyses of density-of-states (DOS), electrostatic potential, charge density differences, and so on. Surprisingly, we find some molecular orbital characteristics which are obviously related to cluster stability. We also identify that the electron delocalization effect is important in the stabilization mechanism at n =3 and 4.

**Results**

We obtained the stable structures for n=3-6 using the structural optimization approach described in the Supporting Information, Part 1. By calculating the frequencies of the structures at the same level, we prove the reliability of the results (see Part 2 of the Supporting Information). We further analyzed the electronic structures in order to explore their structural stability. Firstly, we calculated the DOS of a water ring, and then examined its molecular orbital characteristics.

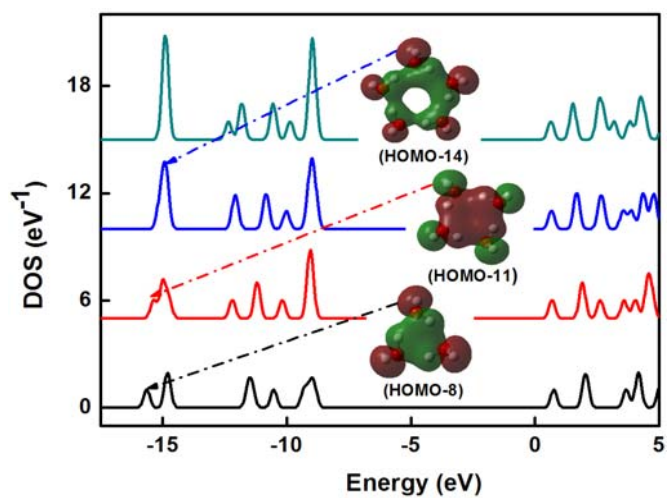

**Figure 1** | DOS of water ring structures. The curves from bottom to top denote, respectively, the DOS for water rings of $3H_2O$, $4H_2O$, $5H_2O$, and $6H_2O$. Black, red, and blue arrows point to the peaks which originate from the corresponding orbitals. The DOS full-width at half-maximum (FWHM) is 0.01 and the isosurface value of the orbital diagram is ±0.02.

Figure 1 clearly shows two weak peaks at around -15.0 eV of the water clusters at n=3 and 4. Further analysis shows that their corresponding molecular orbitals present broadened delocalization areas among the water molecules. More interestingly, the delocalized orbital is formed by the hydrogen atoms involved. Note that a similar electron delocalization of the hydrogen atom as a result of the hyperconjugation effect has been previously reported and shown to be responsible for the stabilization of hydrocarbon structures[22,23]. With the number of water molecules gradually increased, the two weak peaks blend into one. From the corresponding molecular orbital (that is, the blue line of n = 5), it can clearly be seen that there is no electronic distribution of the central area. With an increase in the number of water molecules (n), the molecular orbital tends to be localized.

| $3H_2O$ (HOMO-8) | | | $4H_2O$ (HOMO-11) | | | $5H_2O$ (HOMO-14) | | |
|---|---|---|---|---|---|---|---|---|
| O | H | | O | H | | O | H | |
| 2p | $1s_H$ | 1s | 2p | $1s_H$ | 1s | 2p | $1s_H$ | 1s |
| 76.01% | 10.17% | 13.37% | 76.29% | 9.56% | 13.72% | 76.37% | 9.30% | 13.88% |

**Table 1** | Atomic orbital contribution percentages in the highest occupied molecular orbitals (HOMO)-(8,11,14) corresponding to the characteristic peaks indicated in Figure 2 (DOS) of $3H_2O$, $4H_2O$, and $5H_2O$. p represents the valence orbital contribution of an oxygen atom, $s_H$ the s orbital contribution of a hydrogen atom forming the hydrogen bond, and s the s orbital contribution of a hydrogen atom not forming any hydrogen bond.

To examine the contribution of the oxygen and hydrogen atoms to the molecular orbitals, we conducted a Natural Bond Orbitals (NBO) analysis using the Natural Atomic Orbitals (NAO) approach based on calculated eigenstates. Their contribution percentages are listed in Table 1. The results of the isolated contributions of the water molecular orbitals (see Supporting Information Part 3) indicate that the 2p electron contribution of the oxygen atom is about 74.32% in the hydrogen bond (see the HOMO-2 in Part 3 of the Supporting Information), and the 1s electron contributions of the two hydrogen atoms are about 12.62%. The linear combination of the molecular orbitals of different fragments (that is, different water molecules) gives the molecular orbital shown in Figure 2, where the p orbital component also varies due to the structural differences. From the orbital analysis, the total contribution of the 2p and 1s orbital electrons are basically the same as that of the isolated water molecule.

**Discussion**

This interesting electronic structural feature can also be seen in the electrostatic potential distribution of the water ring (without constraints, which is consistent with that of the free water rings shown in Part 4 of the Supporting Information). It can be seen from Figure 2 that the electron distributes throughout the central region at n= 3 and 4, which is consistent with the DOS result. This also shows that the presence of a stable water ring structure depends on intermolecular hydrogen bonding and electron delocalization in the central region. Starting from n=5, the van der Waals boundary appears in the ring center (where the electron density is less than 0.001). As the number of water molecules increases, the central area (the van der Waals boundary expansion) enlarges, consistent with the change in the DOS.

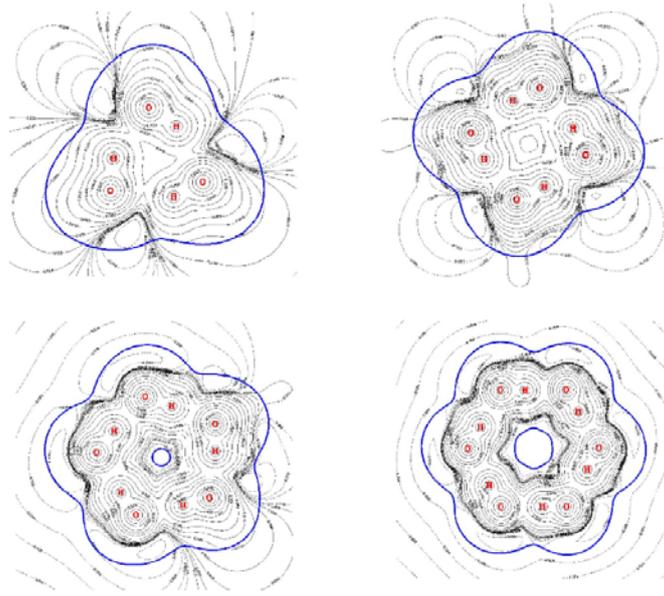

**Figure 2** | Electrostatic potentials of the water ring structures for n=3-6. The blue curve represents a van der Waals border and the red circle an oxygen and hydrogen atom. The van der Waals boundary is at an electron density equal to 0.001 a.u. when the electrostatic potential is less than 0.001 (small enough). Hence, the electron is not so strong that the area outside the van der Waals boundary can be considered as a weak interaction region between electrons. The electrostatic potential is denoted as $\phi$ at a point r arising from M nuclei and the electron density ρ and is calculated[24,25] by

$$\phi(r) = \sum_{a=1}^{M} \frac{z_a}{|R_a - r|} - \int \frac{\rho(r')}{|r' - r|} dr'$$

. Here $R_a$ and $Z_a$ are the position and the charge of nucleus $a$.

The characteristics of orbital delocalization mainly reflect the electronic distribution. For confirmation, we also calculated the charge density differences, defined as ρ=$\rho_{Total}$-Σ$\rho_{Molecule}$ (where $\rho_{Total}$ is the electron density of the water ring and $\rho_{Molecule}$ is the electron density of each water molecule).

From the charge density differences in Figure 3, it is clear that the oxygen atom has acquired an electron and the hydrogen atom has lost one. There is also more charge in the middle region at n =3 and 4 than in other regions, meaning that stability increases in ring forms. From the above analyses of electrostatic potential and charge density difference, we can see that compared to the case of n=5, the electron delocalization for n=3 and 4 extends over the whole region. The delocalized electron density of the middle region decreases with the increase in the number of water molecules. The electrons tend to localize on the water molecules. Such delocalization characteristics involving hydroxide have been observed experimentally in an intermolecular bonding system (Copper crystal surface binding on the dehydrogenated hydroxyquinoline molecule)[26]. This indicates that the stabilization mechanism of the water ring (n=3 and 4) is different from that of other cluster structures (n>4).

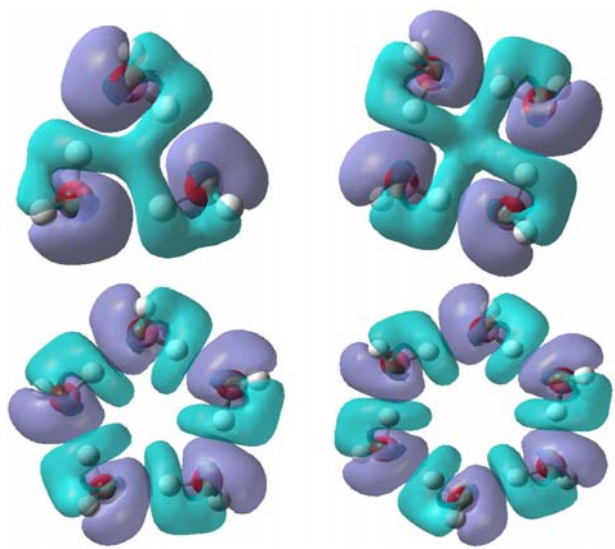

**Figure 3** | Isosurface of charge density difference for water ring structures of n=3-6. The blue region represents the electron donor, and the purple region the electron acceptor. The isosurface value is ±0.0004.

In order to further understand the stabilization mechanism of ISWR with n=3 and 4, we decomposed the more detailed interaction energy, as shown in Figure 4. The average interaction energy is equal to the total energy of the complexes less the energies of all isolated fragments and divided by the number of water molecules. In general, the total interaction energy can be decomposed into two parts; the total energy of the electrostatic and exchange repulsion terms, and the electronic density polarization energy (the correlation between electrons). As shown in Figure 4, the total average interaction energy is significantly affected by the polarization term. The electrostatic and exchange repulsion item decreases monotonically with the increase in n, and the proportion of the electrostatic and exchange repulsion items tends to drop. Previous work indicates that in the system formed by an intermolecular interaction with covalent bonding, high electron density leads to a stronger repulsion term, as seen in polycyclic aromatic hydrocarbons and fullerenes[27]. From Figure 4, for n=3 and 4, the proportion of electrostatic and exchange repulsion is particularly large, comprising up to 33% of the interaction energy for the 3H$_2$O cluster. This also shows that the contribution of the electrostatic and exchange repulsion interaction plays a key role in the formation of small water cluster structures.

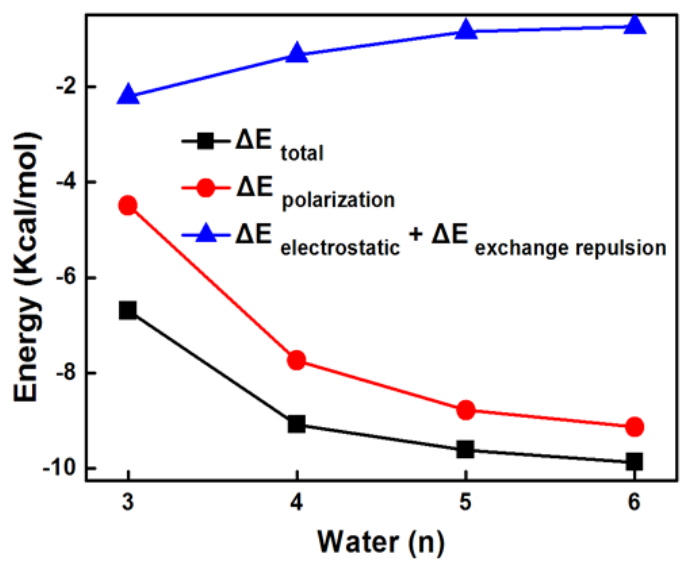

**Figure 4** | Decomposed energies of water ring structures. Black solid lines represent the average total energy variation (average interaction energy), the red solid line the average electronic density polarization component, and the blue solid line the average sum of the electrostatic interaction and exchange repulsion components. The energy decomposition equation used here is[25]:

$$\Delta E_{total} = (\Delta E_{electrostatic} + \Delta E_{exchange\ repulsion}) + \Delta E_{polarization}.$$

Each item in the expression is the average result per water molecule in the water cluster. $\Delta E_{total}$ is the total energy variation; $\Delta E_{electrostatic}$ corresponds to the electrostatic interaction energy, $\Delta E_{exchange\ repulsion}$ the exchange repulsion interaction energy, and the $\Delta E_{polarization}$ the electrostatic density polarization component.

Since the geometric structure of hydrogen bonds can be a good indicator of the stability of the system[9,21], we selected the first of the two hydrogen atoms with the same orientation as the starting point to count all the hydrogen bond lengths of $R_{hb}$ (the distance between the donor hydrogen and acceptor oxygen atoms) and bond angle $\phi$

(hydrogen bond angle O-H…O) in the counterclockwise direction. We note that the two $H_s$ atoms are found in the water ring with an odd number of water molecules in Part 1 of the Supporting Information. There is no hydrogen atom with the same orientation in the water ring structure with an even number of water molecules, and the hydrogen bonds in these rings are all the same. As shown in Figure 5, an increase in the number of water molecules corresponds to a gradual decrease in the average and up-(up and down) hydrogen bond lengths for n= 3-6, consistent with the literature[28] and with a gradual increase in the average and up-(up and down) bond angles. The latter is consistent with previous findings[21]. A detailed analysis can be found in Part 5 of the Supporting Information. More interestingly, similar to the energy decomposition shown in Figure 4, the analysis of hydrogen bond length and bond angle also shows a rapid change, consistent with the characteristic changes in electronic structure in the systems of n=3 and 4. The variation in the NBO and Mulliken charges follow a similar trend. The results are shown in Part 7 of the Supporting Information.

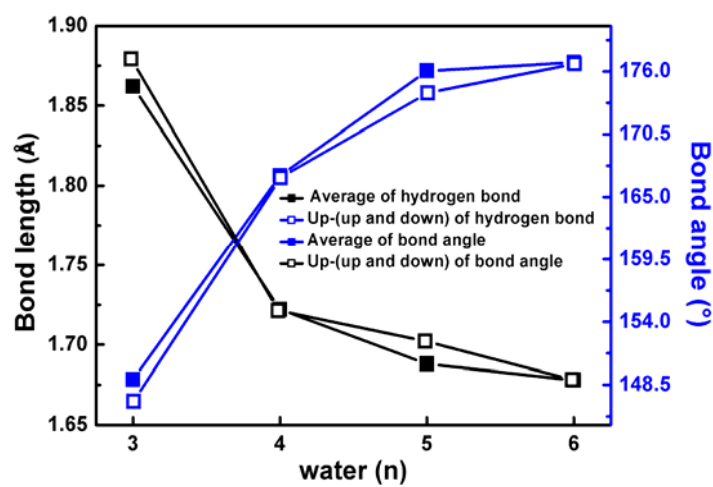

**Figure 5** | The hydrogen bond ($R_{hb}$) and bond angle (O-H…O) measured using the hydrogen atoms of the same orientation in Figure S1 as the starting point and with a counter-clockwise rotation. The black solid square represents the average hydrogen bond length and the black open-square the up-(up and down) hydrogen bond length. There is no up-up (down) hydrogen atom when n is even (odd), so the empty and solid boxes are at the same point. The blue solid square line represents the averaged, and the blue open square the up-(up and down), of the bond angle (O-H…O).

It is worth noting that recent experiments show that hydrogen bonds with covalent Bonds demonstrate similar features[29]. The covalent properties of an 8-hydroxyquinoline molecule assembled on Cu(111) substrate using noncontact atomic fore microscopy (NC-AFM) have also been observed[26]. Our work shows the presence of delocalized electronic distribution characteristics, reflecting covalent characteristics in the most general hydrogen bonding systems of water rings. This result is an important reference point for investigation of the hydrogen bonding nature and structural properties of water clusters.

There are obvious delocalized electronic characteristics in the two smallest water rings (n=3 and 4). In particular, there are obvious differences in DOS, electrostatic potential, charge density difference, and hydrogen bond geometry structure compared with larger rings. Energy decomposition, especially for n=3, reveals that the electrostatic and exchange repulsion makes a substantial contribution to all the average interaction energies. Our findings are an important reference point for exploring the hydrogen bonding nature and structural properties of water clusters.

**Methods**

DFT methods have been shown to provide reliable results in studying the characteristics of water clusters. They have been widely used in previous studies of water clusters[18]. Owing to the inclusion of electron correlation effects, DFT methods have an advantage in describing hydrogen bonding interactions[30,31]. Although the post Hartree-Fock approaches (such as 2nd order Möller-Plesset perturbation theory (MP2)[32] and coupled-cluster theory with single double (triple) excitations (CCSD(T))[33]) can deal accurately with such interactions, they require considerably more computational resources (especially CCSD(T))[34].

Among the various DFT methods, the hybrid Perdew-Burke-Ernzerhof (PBE0) functional[35-37] is optimized to deal with hydrogen bonding interactions in water clusters [3,5], using double-zeta basis sets[15,16,21,38]. As the essence of hydrogen bonding is intermolecular interactions, it is necessary to use polarization and diffuse functions in the basis set [15,39,40]. Therefore, in this study we used a 6-31+G(d, p) basis set for all atoms. All geometric structures were obtained using the PBE0 method. MP2 and M06-2X methods[41] were also performed for the purposes of comparison, with details given in Part 6 of the Supporting Information. All the calculations were performed using the Gaussian 09 package[42].

**Acknowledgements**

The work was supported by the National Science Foundation of China under grant nos. 11374004 and 11004076. Z. W. also acknowledges the assistance of the High Performance Computing Center (HPCC) of Jilin University.

**References**

1. Xantheas, S. S. Ab initio studies of cyclic water clusters $(H_2O)_n$, n=1–6. II. Analysis of many-body interactions. *J. Chem. Phys.* **100**, 7523-7534(1994).

2. Vorgelegt von, M. Sc., Biswajit, S. B. Density-Functional Theory Exchange-Correlation Functionals for Hydrogen Bonds in Water. 2010.

3. Santra, B. et al. On the accuracy of density-functional theory exchange-correlation functionals for H bonds in small water clusters. II. The water hexamer and van der Waals interactions. *J. Chem. Phys.* **129**, 194111(2008).

4. Xantheas, S. S. Cooperativity and hydrogen bonding network in water clusters. *Chem. Phys.* **258**, 225-231(2000).

5. Dahlke, E. E. & Truhlar, D. G. Improved Density Functionals for Water. *J. Phys. Chem. B* **109**, 15677-15683(2005).

6. Todorova, T., Seitsonen, A. P., Hutter, J., Kuo, I-F., Mundy, C. J. Molecular Dynamics Simulation of Liquid Water Hybrid Density Functionals. *J. Phys. Chem. B* **110**, 3685-3691(2006).

7. Sadlej, J., Buch, V., Kazimirski, J., K. Theoretical Study of Structure and Spectra of Cage Clusters $(H_2O)_n$, n=7-10. *J. Phys. Chem. A* **103**, 4933-4947 (1999).

8. Maheshwary, S., Patel, N., Sathyamurthy, N., Kulkarni, A., D., & Gadre, S., R. Structure and Stability of Water Clusters $(H_2O)_n$, n=8-20 An Ab Initio Investigation. *J. Phys. Chem. A* **105**, 10525-10537(2001).

9. Xantheas, S. S. Ab initio studies of cyclic water clusters $(H_2O)_n$, n=1–6. III. Comparison of density functional with MP2 results. *J. Chem. Phys.* **102**, 4505-4516(1995).

10. Pugliano, N. Saykally, R. J. Measurement of quantum tunneling between chiral isomers of the cyclic water trimer. *Science, New Series* **257**, 5078(1992).

11. Gregory, J. K. The Water Dipole Moment in Water Clusters. *Science* **275**, 814-817(1997).

12. Huisken, F., Kaloudis, M. & Kulcke, A. Infrared spectroscopy of small size-selected water clusters. *J. Chem. Phys.* **104**, 17(1996).

13. Saykally, R. J. & Wales, D. J. Pinning down the water hexamer. *Science* **336**, 814-815(2012).

14. Liu, K. et al. Characterization of a cage form of the water hexamer. Letters TO Nature **381**, 501–503(1996).

15. Shield, R. M., Tomelso, B., Archer, K. A., Morrell, T. E., & Shieds, G. C. Accurate Predictions of Water Cluster Formation, (H2O)n2-10. *J. Phys. Chem. A* 114, 11725-11737(2010).

16. Losada, M. and Leutwyler, S. Water hexamer clusters: Structures, energies, and predicted mid-infrared spectra. *J. Chem. Phys.* **117**, 2003-2015(2002).

17. Temelso, B., Archer, K. A. & Shields, G. C. Benchmark structures and binding energies of small water clusters with anharmonicity corrections. *J. Phys. Chem. A* **115**, 12034-12046(2011).

18. Su, J. T., Xu, X. & Goddard III, W. A. Accurate Energies and Structures for Large Water Clusters Using the X3LYP Hybrid Density Functional. *J. Phys. Chem. A* **108**, 10518-10526(2004).

19. Niu, Y. L. et al. Quantum chemical calculation of intramolecular vibrational redistribution and vibrational energy transfer of water clusters. *Chem. Phys. Lett.* **586**, 153-158(2013).

20. Hirunsit, P., Balbuena, P. B. Effects of Confinement on Small Water Clusters Structure and Proton Transport. *J. Phys. Chem. A* **111**, 10722-10731(2007).

21. Xantheas, S. S., Jr, T. H. D. Ab initio studies of cyclic water clusters (H2O)n, n=1–6. I. Optimal structures and vibrational spectra. J. Chem. Phys. **99**, 8874-8792(1993).

22. Meng, Y.A. et al. Hyperconjugation Effect On the Structural Stability of a Tert-Butyl and Its Derived $C_4H_n$(n = 4–10) ISOMERS. *J. Theor. & Comput. Chem.* **11**, 1217-1225(2012).

23. Goodman, L., Sauers, R. R. 1-Fluoropropane. Torsional Potential Surface. *J. Chem. Theory Comput.* **I**, 1185-1192(2005)

24. Jakobsen, S., Kristensen, K. & Jensen, F. Electrostatic Potential of Insulin: Exploring the Limitations of Density Functional Theory and Force Field Methods. *J. Chem. Theory Comput.* **9**, 3978-3985(2013).

25. Lu, T., Chen, F. W. Multiwfn: A Multifunctional Wavefunction Analyzer. J. Comp. Chem. **33**, 580-592(2012).

26. Zhang, J. et al. Real-Space Identification of Intermolecular Bonding with Atomic Force Microscopy. Science **342**, 611(2013).

27. De Oteyza, D. G.,et al. Direct imaging of covalent bond structure in single-molecule chemical reactions. *Science* **340**, 1434(2013).

28. Sun, C.Q., et al.,Density, Elasticity, and Stability Anomalies of Water Molecules With Fewer-Than-Four Neighbors. *J. Phys. Chem. Lett.* **4**,15(2013).

29. Espinosa, E., Lecomte, C. L. & Molins, E. Experimental electron density overlapping in hydrogen bonds topology vs. energetics. *Chem. Phys. Lett.* **300**, 745-748(1999).

30. Kolb, B. & Thonhauser, T. van der Waals density functional study of energetic, structural, and vibrational properties of small water clusters and ice $I_h$. *Phys. Rev. B* **84**, 045116(2011).

31. Li, F. *et. al.* What is the best density functional to describe water clusters: evaluation of widely used density functionals with various basis sets for $(H_2O)_n$ (n = 1–10). *Theor. Chem. Acc.* **130**, 341-352(2011).

32. Mo/ller, C. & Plesset, M. S. *Phys. Rev.* **46**, 618(1934).

33. Pople, J. A.,Head-Gordon, M. & Raghavachari, K. *J. Chem. Phys.* **87**,5968 (1987).

34. Tsuzuki, S. T. & Lüthi, H.P. Interaction energies of van der Waals and hydrogen bonded systems calculated using density functional theory: Assessing the PW91 model. *J. Chem. Phys.* **114** ,3949(2001).

35. Adamo, C. & Barone, V. Toward reliable density functional methods without adjustable parameters: The PBE0 model. *J. Chem. Phys.* **110**, 6158-6170 (1999).

36. Perdew, J. P., Burke, K. & Ernzerhof, M. Generalized Gradient Approximation Made Simple. *Phys. Rev. Lett.* **77**, 865-3868(1996).

37. Ernzerhof, M., Scuseria, G. E. Assessment of the Perdew-Burke-Ernzerhof exchange-correlatiom functional. *J. Chen. Phys.* **110**, 5029(1999).

38. Hammond, J. R., Govind, N., Kowalski, K., Autschbach, J., & Xantheas, S. S. Accurate dipole polarizabilities for water clusters n=2-12 at the coupled-cluster level of theory and benchmarking of various density functionals. *J. Chem. Phys.* **131**, 214103(2009).

39. Dahlke, E. E., Olson, R. M. Assessment of the Accuracy of Density Functionals for Prediction of Relative Energies and Geometries of Low-Lying Isomers of Water Hexamers. *J. Phys. Chem. A* **112**, 17(2008).

40. Ghadar, Y., Clark, A. E. Coupled-cluster, Moller Plesset (MP2), density fitted local MP2, and density functional theory examination of the energetic and structural features of hydrophobic solvation: water and pentane. *J. Chem. Phys.* **136**, 5(2012).

41. Zhao, Y., Truhlar, D. G. The M06 suite of density functionals for main group thermochemistry, thermochemical kinetics, noncovalent interactions,excited states, and transition elements: two new functionals and systematic testing of four M06-class functionals and 12 other functionals. *Theor. Chem. Acc.* **120**, 215-241(2008).

42. Frisch, M. J. et al. Gaussian 09, revision A. 01, Gaussian, Inc., Wallingford CT, 2009.

# Supplementary Information

**Contents**

**Part 1. The ISWR initial structures**

To prepare the calculations and analysis of the ISWR, we first built the ISWR initial structures containing virtual atoms (see Figure S1). The exo-ring hydrogen of the even number of water molecules lie in up-down (unusual orientation of hydrogen) position with respect to its main stable structure, while the ones with odd number of water molecules exist at up-up (orientation of hydrogen) position with a stable structure.

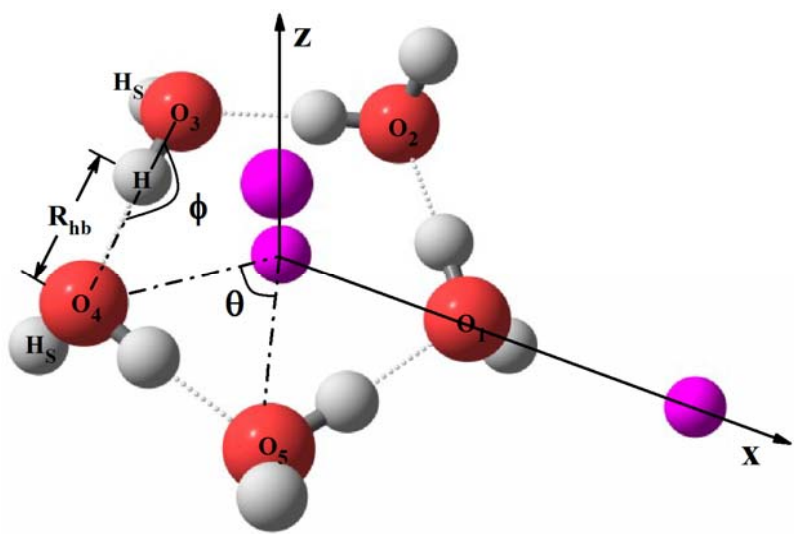

**Figure S1** | A schematic diagram of a typical ISWR structure (oxygen in red, hydrogen in white, and hydrogen bonds in grey dotted line).

To unify the numbering, we call the hydrogen bond in the water molecular containing $O_3$ pointing to the $O_4$ as the first hydrogen bond and the number of the other hydrogen bond in the ring increase in counter-clockwise direction. In this work, all the statistics are done with this definition. $R_{hb}$ represents the hydrogen bond distance (between the donor hydrogen atom and acceptor oxygen atom), $\phi$ the bond angle ($\angle$(O—H…O)), $\theta$ the central angle formed by the two oxygen atoms and one virtual atom, $H_S$ the hydrogen atoms with the same orientation, H hydrogen atom forming hydrogen bond, and $O_1$-$O_5$ the water molecules numbered beginning from the one at the X axis. Based on the position relations of three virtual atoms (pink) the reference axes are established.

**Part 2. Water cluster vibrational characteristics in the IR and Raman spectra**

Infrared and Raman spectroscopies are widely used to describe the structural properties of water clusters. We calculated the vibration spectra of stable structures of water rings with n=3-6 and obtained their Raman and Infrared spectra. The corresponding peak frequencies of Raman spectra are 3569.8 cm$^{-1}$, 3498.7 cm$^{-1}$, 3255.00 cm$^{-1}$, and 3234.9 cm$^{-1}$ for n=3-6, respectively, as shown in Fig 2Sa. These peaks have the same vibration mode, i.e. the overall hydrogen bond stretching mode. We found that the highest peak presents red shifts with the increase of the number of water molecules (n) when n≤6, indicating a trend of bonding strength weakening.

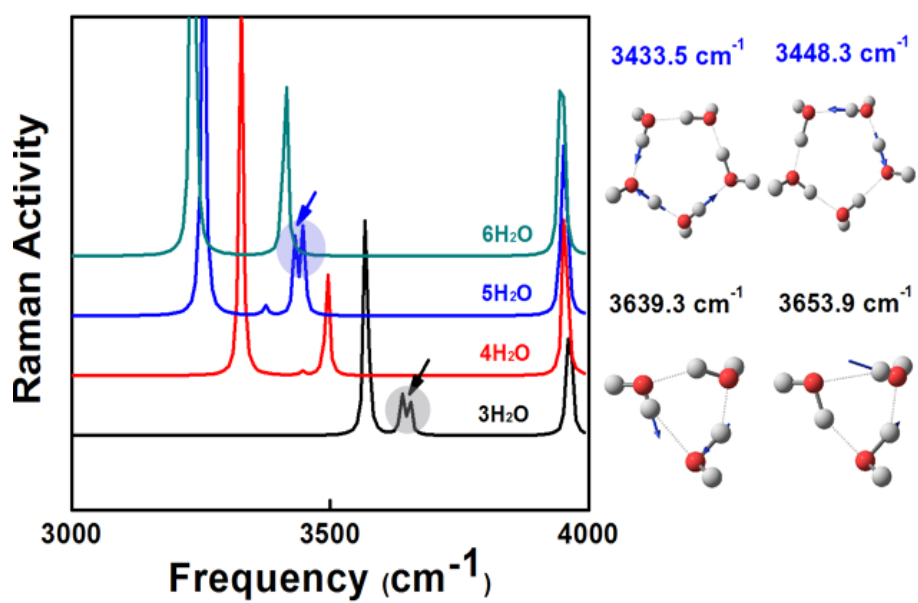

**Figure 2Sa** | The solid lines from the bottom to top respectively denote the Raman spectra for water ring 3H$_2$O, 4H$_2$O, 5H$_2$O and 6H$_2$O. The shadow areas indicated by the blue and black arrows represent the characteristic peaks of two kinds of water ring structures. The right panel illustrates the four spectral vibrational modes of the water ring for 3H$_2$O and 5H$_2$O, and the according modes can be found in the left panel.

In Figure 2Sa, the shaded areas of the peaks to which the black and blue arrows point correspond to the characteristic water ring vibration modes for n= 3 and 5 (see Figure 2Sa for the vibration mode diagram on the right panel). The peaks of black curve of 3639.3 cm$^{-1}$ and blue curve of 3433.5 cm$^{-1}$ respectively correspond to the hydrogen bonds stretching which are away from the water molecules having the same orientations; the peaks on black curve at 3653.9 cm$^{-1}$ and blue curve at 3448.3 cm$^{-1}$ respectively are hydrogen bond stretching modes which are from the water molecules having same orientation. Compared with the Raman spectroscopy, the infrared spectroscopy also show that some qualitative trends.

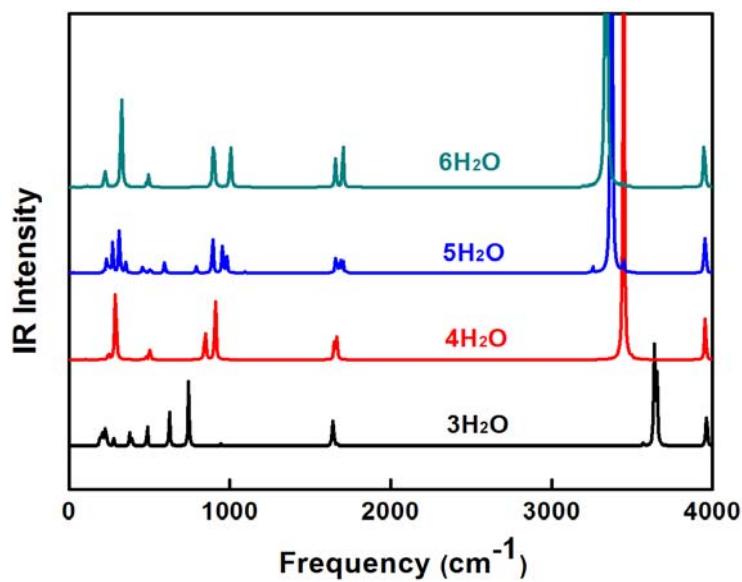

**Figure 2Sb** | The solid curves from bottom to top in the figure are respective the IR spectra for water ring $3H_2O$, $4H_2O$, $5H_2O$, $6H_2O$.

In Figure 2Sb, the peaks of IR spectra can be classified into three regions: (1) the vibration modes within 1100 cm$^{-1}$ correspond to the relative movements of between water molecules and the oxygen-hydrogen bond of water intramolecular asymmetric twist; (2) in the range of 1000-2000 cm$^{-1}$, the vibration mode is the symmetrical swing of the hydrogen and oxygen atoms; and (3) the main vibrational mode in the frequency higher than 3000 cm$^{-1}$ is the stretching modes between the water intermolecular and intramolecular of the hydrogen bond. The highest peaks of the water ring structures are at the frequencies of 3639.3 cm$^{-1}$, 3447.3 cm$^{-1}$, 3366.0 cm$^{-1}$, 3337.9 cm$^{-1}$, respectively. They have the same vibrational mode corresponding to the vibration of the hydrogen bonds. We found that the highest peaks show redshifts, with the increase in the number of water molecules to up to 6.

**Part 3. Molecule orbital contribution of an isolate water**

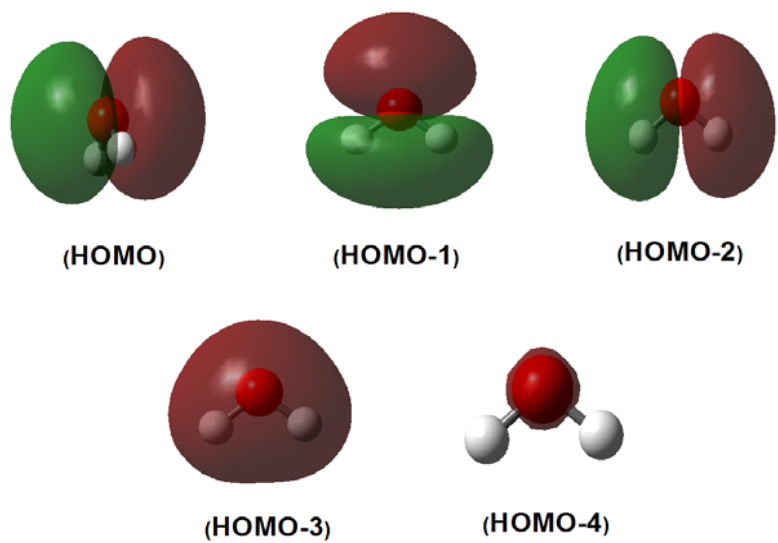

(HOMO)  (HOMO-1)  (HOMO-2)

(HOMO-3)  (HOMO-4)

**Figure S3** | The molecule orbital of the isolate water molecule. We can see clearly that the molecular orbital of the Figure 1 is HOMO-2 which is a linear combination of hydrogen and oxygen atomic orbitals. The 2p orbital contribution of HOMO-2 of the isolate water molecule is about 74.32%, and the 1s orbital contributions are about 12.62% and 12.62%, respectively.

**Part 4. Electrostatic potential of water rings in Cartesian coordinates**

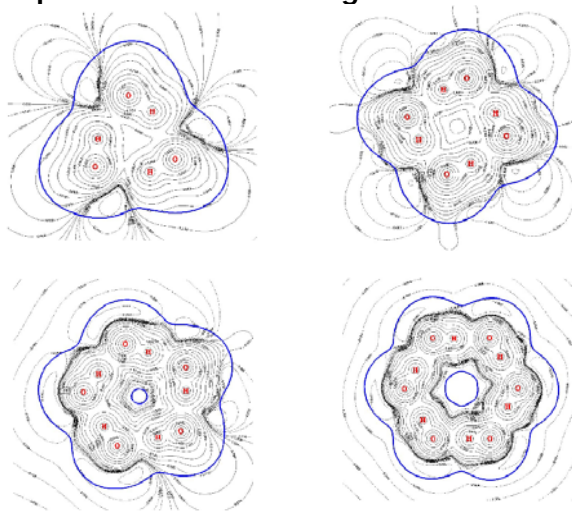

**Figure 4S** | Electrostatic potentials of the water clusters with n=3-6. Blue represents a van der Waals boundary and red an oxygen atom and a hydrogen atom. The van der Waals boundary is at the value of electrostatic potential of 0.001. The electrostatic potential less than 0.001 (small enough). Therefore, the area outside the boundary of the van der Waals can be considered as the weak interaction between electrons.

## Part 5. Analysis of hydrogen bond lengths and bond angles

**Table S1** | Hydrogen bond length $R_{oh}$ (Å)

| | | | | | | |
|---|---|---|---|---|---|---|
| 3H$_2$O | 1.879 | 1.853 | 1.853 | | | |
| 4H$_2$O | 1.721 | 1.721 | 1.721 | 1.723 | | |
| 5H$_2$O | 1.702 | 1.685 | 1.684 | 1.684 | 1.686 | |
| 6H$_2$O | 1.678 | 1.678 | 1.678 | 1.677 | 1.677 | 1.677 |

**Table S2** | Bond angle (O-H…O) $\phi$ (°)

| | | | | | | |
|---|---|---|---|---|---|---|
| 3H$_2$O | 147.2 | 149.9 | 149.8 | | | |
| 4H$_2$O | 166.8 | 166.9 | 166.8 | 167.1 | | |
| 5H$_2$O | 174.2 | 177.0 | 176.2 | 176.4 | 176.5 | |
| 6H$_2$O | 176.8 | 176.8 | 176.8 | 176.8 | 176.8 | 176.8 |

In the table S1 and table S2, the items in the vertical column and diagonal terms (boxes enclosed with dashed lines) increase with the increase in the number of water molecules. The hydrogen bond length decreases for n = 3-6 are consistent with previous report [1,2]. As we can see from the vertical column and diagonal terms in table S2, the bond angle $\phi$ (OH…O) increases gradually with the increase of the water molecules for n=3-6 [1,2].

## Part 6. Energy decomposition based on MP2 and M062X calculations

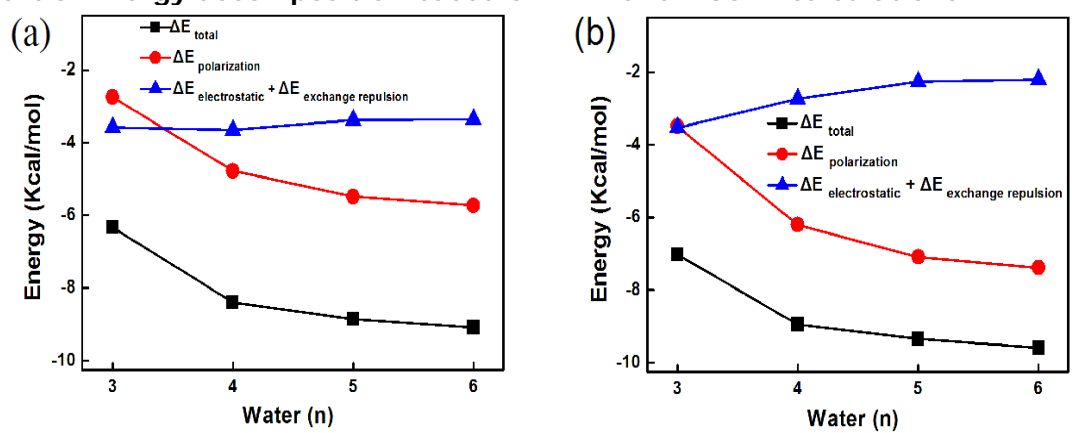

**Figure 6S** | Decomposed energies of water ring structures. Figures a and b respectively based on MP2 and M062X calculations. Black solid lines represent the average total energy variation (average interaction energy), red solid lines the average electronic density polarization energy, and the blue solid lines the sum of the average electrostatic interaction component and exchange repulsion component.

From the MP2 and M062X results in Figure 6S, we can clearly see that with the increase of the number of water molecules (n), the overall average interaction energy decrease, due to the influence of polarization, with the same slow monotonic decrease in polarization, and the electrostatic and exchange repulsion components.

**Part 7. Analysis of the Mulliken and NBO charge of oxygen atoms in water clusters**

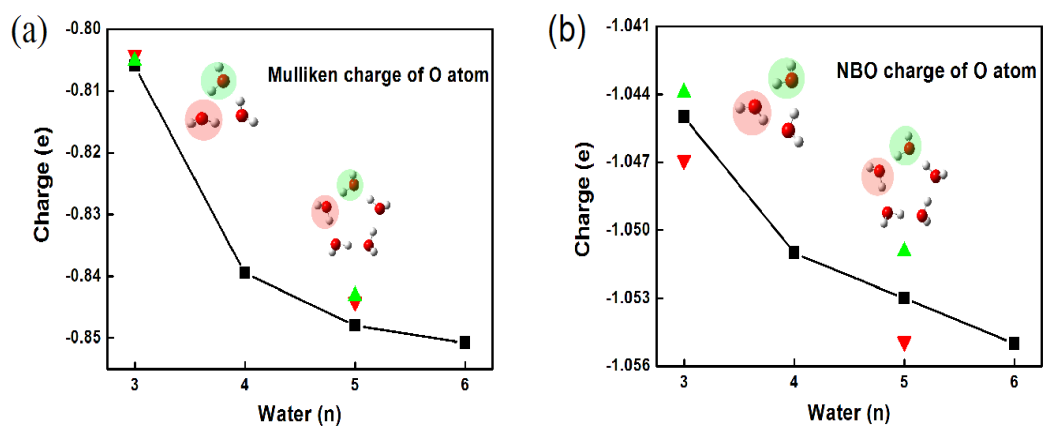

**Figure 7S** | (a, b) represent the Mulliken charge and NBO charge of oxygen atom.

Black boxes represent the average Mulliken charge (or NBO) charge of oxygen atoms at n = (4,6) or average Mulliken charge (or NBO charge) for removal of the oxygen atoms of the same orientation at n = 3 and 5. Red and green areas respectively represent Mulliken charge (or NBO charge) of oxygen atom in the same orientation. Figure 7S (a, b) shows the Mulliken and NBO charge of oxygen atom, which present a similar trend for n=3-6: the Mulliken or NBO charge of oxygen atom show an obvious charge change at n=3 and 4.

**Part 8. References for Supplementary Information**

S1. Xantheas, S. S. Ab initio studies of cyclic water clusters $(H_2O)_n$, n=1–6. III. Comparison of density functional with MP2 results. *J Chem. Phys.* **102**, 4505-4516(1995).

S2. Xantheas, S. S., Jr, T. H. D. Ab initio studies of cyclic water clusters $(H_2O)_n$, n=1–6. I. Optimal structures and vibrational spectra. *J. Chem. Phys.* **99**, 8874-8792(1993).